\documentclass[epj]{svjour}
\usepackage{epsfig,array}
\usepackage{multirow}
\usepackage{amsmath, amssymb,epsfig, pifont}
\usepackage{array, tabularx, multirow}
\usepackage{graphicx}
\def\nn{\nonumber \\}

\def\fun#1#2{\lower3.6pt\vbox{\baselineskip0pt\lineskip.9pt
\ialign{$\mathsurround=0pt#1\hfil##\hfil$\crcr#2\crcr\sim\crcr}}}

\newcommand{\be}{\begin{eqnarray}}
\newcommand{\ee}{\end{eqnarray}}

\def\nn{\nonumber \\}

\def\fun#1#2{\lower3.6pt\vbox{\baselineskip0pt\lineskip.9pt
\ialign{$\mathsurround=0pt#1\hfil##\hfil$\crcr#2\crcr\sim\crcr}}}

\begin{document}

\title{Polarization degrees of freedom in near-threshold photoproduction of
$\mathbf \omega$ mesons in the $\mathbf{\pi^0\gamma}$ decay channel}

\author{
A.V.~Sarantsev\inst{1,2}, A.V.~Anisovich\inst{1,2},  V.A.~Nikonov
\inst{1,2} and H.~Schmieden\inst{3} }

\institute{ Helmholtz--Institut f\"ur Strahlen-- und Kernphysik,
Universit\"at Bonn, Germany \and Petersburg Nuclear Physics
Institute, Gatchina, Russia \and Physikalisches Institut,
Universit\"at Bonn, Germany}

\date{\today}

\abstract{We study polarization variables in the photoproduction of
$\omega$-mesons with subsequent $\omega \rightarrow \gamma \pi^0$
decay. Single and double polarization observables are calculated as
a function of different final-state angles. Reaction models include
pomeron (natural parity) and $\pi^0$ (unnatural parity) exchange in
the $t$-channel. In addition, the contribution of s-channel
resonances is considered. The sensitivity of the polarization
observables to the reaction dynamics is discussed.}

 \PACS{
     {11.80.Et}{Partial-wave analysis}
\and {13.30.-a}{Decays of baryons} \and {13.60.Le}{Meson
production}}

\authorrunning{A.V.~Sarantsev {\it et al.}}
\titlerunning{Polarization degrees of freedom in near-threshold
              photoproduction of $\omega$ mesons}

\mail{andsar@iskp.uni-bonn.de}

\maketitle

\section{Introduction}

The hadron mass spectrum and hadronic decay modes provide important
information towards the understanding of strong interactions at low
and intermediate energies. While in the meson sector the number of
observed states roughly corresponds to the prediction of the quark
model, in the baryon sector the expected spectrum of such models
\cite{Isgur:1995ei,Capstick:bm,Riska,Metsch} is much richer than
observed \cite{pdg}. The possibility is discussed that the {\it
missing resonances} may have escaped observation in pion induced
processes due to their small $N\pi$ coupling
\cite{Capstick:bm,IK77,KI80}. The {\em photo}production of mesons
offers a means to get a hand on the existence of the missing states,
in particular in non-pionic final states.

The photoproduction of $\omega$ mesons off the proton is well suited
to investigate this issue. It benefits from the fact that the
$\omega$ threshold is in the higher lying third resonance region of
the nucleon. The narrow width of 8 MeV enables a clean detection of
the $\omega$ over background. In addition, the $\omega$ is isoscalar
($I=0$); hence a s-channel process will only connect $N^*$ ($I=1/2$)
states with the nucleon ground state, but no $\Delta^*$ with
$I=3/2$. This provides a great simplification to the complexity of
the contributing excitation spectrum.

Several recent experiments \cite{Barth03,Ambrozewicz04,Ajaka06} and
coupled channel analyses \cite{Penner02,Shklyar05} reported evidence
for resonance contributions close to $\omega$ production threshold.
Usually the $\omega$ is detected through the charged decay $\omega
\rightarrow \pi^+\pi^-\pi^0$. The neutral $\omega \rightarrow
\pi^0\gamma$ decay provides complementary information \cite{ZAC05}.
Recently, first beam asymmetries have been measured using this
channel \cite{HS07}.

In order to disentangle the reaction mechanism and characterize
individual resonance contributions, the measurement of polarization
observables is indispensable. A completely unambiguous decomposition
of the reaction amplitudes would require at least 23 independent
observables to be measured, including double and triple polarization
observables.

It is the scope of the present paper to investigate single and
double polarization observables in the reaction $\gamma p
\rightarrow \omega p \rightarrow \pi^0 \gamma p$. The sensitivity of
the observables to resonance contributions to $\omega$
photoproduction is investigated.

The paper is organized as follows. Cross section and polarization
observables are first discussed on the basis of the reaction
dynamics. The influence of the reaction mechanism on the observables
is then investigated in section 3, in particular the sensitivity to
$s$-channel resonance contributions. The main results are finally
summarized.

\section{Cross sections, angular distributions and polarization
variables}

The differential cross section for production of two or more
particles has the general form
\be
d\sigma=\frac{(2\pi)^4
|A|^2}{4\sqrt{(k_1k_2)^2-m_1^2m_2^2}} d\Phi_n(k_1+k_2,q_1,\ldots,q_n)
\ee
where $k_1$ and $k_2$ are the momenta of the initial particles
$k_i^2=m_i^2$ (nucleon and $\gamma$ in the case of photoproduction)
and $q_i$ denote the momenta of the final state particles. $d\Phi_n$
is the n-body phase volume given by
\be
d\Phi_n(k_1+k_2,q_1,\ldots,q_n)=
\qquad\qquad\qquad\qquad\qquad\nonumber\\
\qquad\qquad\ \delta^4(k_1+k_2-
\sum\limits_{i=1}^n q_i)
\prod\limits_{i=1}^n \frac{d^3q_i}{(2\pi)^3 2q_{0i}}\;.
\label{phv}
\ee
The amplitude for photoproduction of a single (pseudo--) scalar
meson is written in the form
\be
A=\varepsilon_\mu \bar u(q_1) A_{\mu} u(k_1) \;,
\ee
where $\varepsilon_\mu$ is the photon polarization vector and $\bar
u(q_1)$ and $u(k_1)$ are bispinors of the final and initial nucleons
with momenta $q_1$ and $k_1$, respectively. When the polarizations
of photon and nucleon  are not measured the squared amplitudes are
averaged over orthogonal polarization vectors of the initial
particles and summed over polarization vectors of the final
particles:
\be
\label{squaredA}
|A|^2\!=\!\frac 14 \sum\limits_{\alpha i j} \varepsilon^\alpha_\mu
\varepsilon^{\alpha *}_\nu \bar u^i(q_1) A_\mu u^j(k_1) \bar
u^j(k_1) A^{tr}_\nu u^i(q_1),
\ee
Here $A^{tr}$ is the hermitian conjugated amplitude. The index
$\alpha$ describes orthogonal polarization vectors of the photon and
$i,j$ describe the two orthogonal polarizations of the initial and
final bispinors.

The amplitude for the photoproduction of vector me\-sons has the
structure
\be
A=\epsilon_\mu \bar u(q_1) A_{\mu\nu} u(k_1) \omega_\nu\,,
\ee
where $\omega_\nu$ is the vector-meson polarization vector. The
squared amplitude for the unpolarized case is given by
\be
|A|^2\!=\! \! \sum\limits_{\alpha i j}
\frac{\varepsilon^{\alpha}_\mu \varepsilon^{\alpha*}_\nu}{4} \bar
u^i(q_1) A_{\mu\tau} u^j(k_1) \bar u^j(k_1) A^{tr}_{\nu\xi} u^i(q_1)
\omega_\tau \omega^*_\xi,~~
\label{a2_om}
\ee

For the $\omega$ meson decaying into three pions, the direction of
the $\omega$ polarization vector is defined by
\be
\omega_\xi=\varepsilon_{\xi\alpha\beta\gamma} q_{2\alpha} q_{3\beta}
q_{4\gamma}
\ee
where $\varepsilon_{\nu\alpha\beta\gamma}$ is the totally
antisymmetrical tensor and $q_i$ are the momenta of the final--state
pions. For the $\pi^0\gamma$ decay, the $\omega$ polarization vector
can be written as
\be
\omega_\xi=\frac{1}{M_\omega}\varepsilon_{\xi\alpha\beta\nu}
q_\alpha p_\beta \tilde \epsilon_\nu\equiv
\frac{1}{M_\omega}\varepsilon_{\xi q p \nu} \tilde \epsilon_\nu
\ee
where $\tilde \epsilon_\nu$ is the polarization vector of the final
photon, $q$ is the pion 4-vector ($q\equiv q_2$) and $p$ is the
4-vector of the $\omega$--meson ($p=q_2+q_3$ where $q_3$ is the
photon 4-vector). The factor $1/M_\omega$ ($M_\omega^2\equiv p^2$)
is introduced to avoid the divergence of the amplitude at large
energies.

If the polarization of the final photon is not measured, the omega
polarization vector can not be completely determined.  However its
direction should be orthogonal to the plane which is defined by the
4-vectors of the $\omega$ and $\pi$ mesons. The product of the
$\omega$ polarization vectors summed over the polarizations of the
final photons yields:
\be
\omega_\tau\omega^*_\xi= -\frac{1}{M_\omega^2}\varepsilon_{\tau\mu q
p}\, \varepsilon_{\xi \nu q p} \,\,\tilde g^{\perp\perp}_{\mu\nu}\,
\label{omega_ten}
\ee
Here $\tilde g^{\perp\perp}_{\mu\nu}$ is a tensor orthogonal to the
momentum of the photon (as any photon polarization vector) and to
the momentum of one of the another on-shell-particle in the vertex
(gauge invariance):
\be
-\sum\limits_{\alpha} \tilde \varepsilon^{\alpha}_\mu \tilde
\varepsilon^{\alpha\,*}_\nu=\tilde g_{\mu\nu}^{\perp\perp}=
g_{\mu\nu}-\frac{p_\mu p_\nu}{M_\omega^2}-\frac{ q^\perp_\mu
q^\perp_\nu}{q_\perp^2}\,.
\label{fin_gamma_ten}
\ee
where
\be
q^\perp_\mu=q_\nu \tilde g^\perp_{\mu\nu},\qquad \tilde
g^\perp_{\mu\nu}=\left ( g_{\mu\nu}-\frac{p_\mu p_\nu}{M_\omega^2}
\right )\,.
\ee
and $q_\perp^2=q^\perp_\mu q^{\perp}_\mu$.
 The tensor $\omega_\tau\omega^*_\xi$, defined in
eq.(\ref{omega_ten}), has the same structure as the tensor $\tilde
g^{\perp\perp}_{\mu\nu}$:
\be
\omega_\tau\omega^*_\xi= -\left (q_\perp^2\tilde g_{\tau\xi}^\perp
-q^\perp_\tau q^\perp_\xi\right )
\label{omega_fin}
\ee

For unpolarized photons the product of polarization vectors summed
over polarizations gives:
\be
-\sum\limits_{\alpha} \varepsilon^{\alpha}_\mu \varepsilon^{\alpha\,
*}_\nu=g_{\mu\nu}^{\perp\perp}= g_{\mu\nu}-\frac{P_\mu
P_\nu}{P^2}-\frac{ k^\perp_\mu k^\perp_\nu}{k_\perp^2}
\ee
where  $P=k_1+k_2$ is the total momentum of the $\gamma N$ system,
$k_1$ is the momentum of the initial nucleon and $k_2$ is the photon
momentum. Then
\be
k^\perp_\mu\!=\!\frac 12 (k_1\!-k_2)_\nu g^\perp_{\mu\nu}\!=\! \frac
12 (k_1\!-k_2)_\nu \left ( g_{\mu\nu}-\frac{P_\mu P_\nu}{P^2} \right
).~~
\ee

In the center-of-mass of the $\gamma N$ reaction and with the photon
momentum directed along the $z$-axis the tensor
$g^{\perp\perp}_{\mu\nu}$ has the form:
\be
g^{\perp\perp}_{\mu\nu} =  \left (
\begin{array}{cccc}
0 & 0 & 0 &0 \\
0 &-1 & 0 &0 \\
0 & 0 &-1 &0 \\
0 & 0 & 0 &0
\end{array}
\right )\;.
\ee

The bispinors of fermions with momentum $k_1$ summed over
polarizations are:
\be
\sum\limits_{j} u^j(k_1)\bar u^j(k_1)=m+\hat k_1\,,\qquad \hat
k_1\equiv k_{1\mu}\gamma_\mu
\ee
and therefore the squared amplitude (\ref{a2_om}) can be rewritten
as:
\be
|A|^2= \frac 14 g_{\mu\nu}^{\perp\perp} Tr\left [(m+\hat k_1)
A_{\mu\tau} (m+\hat q_1) A^{tr}_{\nu\xi} \right ]\times \nn
(q_\perp^2 \tilde g_{\tau\xi}^\perp -q^\perp_\tau q^\perp_\xi )\;.
\label{am2_fin}
\ee
It is stressed that the tensor $g_{\mu\nu}^{\perp\perp}$ is
orthogonal to the total momentum of the $\gamma N$ system, while
$\tilde g^\perp_{\tau\xi}$ and $q^\perp_{\xi}$ are orthogonal to the
momentum of the $\omega$-meson.

\subsection{Photon polarizations}

When the initial photon is linearly polarized along the $y$-axis,
the polarization vector in the laboratory and c.m. system is
$\varepsilon_\mu=(0,0,1,0)$ and we do not need to average over the
polarizations. In eq.(\ref{am2_fin}) then
\be
\frac 12 g^{\perp\perp}_{\mu\nu}\qquad\to\qquad
\left (
\begin{array}{cccc}
0 & 0 & 0 &0 \\
0 & 0 & 0 &0 \\
0 & 0 &-1 &0 \\
0 & 0 & 0 &0
\end{array}
\right )
\ee

For a circularly polarized beam with helicity $\pm 1$,
$\varepsilon_\mu=\mp\frac{1}{\sqrt 2}(0,1,\pm i,0)$. For helicity
$+1$ the density matrix is
\be \frac 12
g^{\perp\perp}_{\mu\nu}\qquad\to\qquad \frac 12 \left (
\begin{array}{cccc}
0 & 0 & 0 &0 \\
0 & -1 & -i &0 \\
0 & i  & -1 &0 \\
0 & 0 & 0 &0
\end{array}
\right )\,.
\ee
Helicity $-1$ differs in the sign of the nondiagonal elements.

The photon beam is assumed to have either linear or circular
polarization. In general, the degree of polarization is not 100\%.
Linear polarization is in the range $0 < P_{-}^{\gamma} < +1$; the
degree of circular polarization $P_{\odot}^{\gamma}$ is between $-1
< P_{\odot}^{\gamma} < +1$.

\subsection{Polarized target and recoil polarization}

For a polarized target, the density matrix of the nucleon propagator
$(m+\hat k_1)$ must be changed to a polarization density matrix:
\be
m+\hat k_1\to (m+\hat k_1)(1+\gamma_5\hat P^t) \;.
\ee
where the 4-vector $P^t$ is the polarization vector of the target
baryon. Assuming 100\% polarization:
\be
P^t=(P^t_0;\vec P^t)\qquad (P^t)^2=-1,\qquad k_1\cdot P^t=0\,.
\ee
If the baryon momentum is directed along the $z$ axis
\be
k_1&=&(\sqrt{m_p^2+k_z^2},0,0,k_z)\,,
\ee
then the longitudinal polarization vector is equal to:
\be
P^t&=&(\frac{k_z}{m_p};0,0,\frac{\sqrt{m_p^2+k_z^2}}{m_p})\,.
\ee

The transverse polarization the polarization vector has only $x$
and/or $y$ components.

If the polarization $P^r$ of the final--state baryon
$P^r=(P^r_0,\vec P^r)$ is measured, the density matrix of the
propagator $(m+\hat q_1)$ is substituted by
\be
m+\hat q_1\to (m+\hat q_1)(1+\gamma_5\hat P^r)\;,
\ee
$P^r$ satisfies $(P^r)^2=-1,\ q_1\cdot P^r=0$.

\subsection{Polarization variables for pseudoscalar meson
photoproduction}

For single pseudoscalar meson photoproduction the description
follows a paper by Chiang, Yang, Tiator and Drechsel \cite{maid}.
The polarized and unpolarized differential cross sections (which we
define for the sake of simplicity as $\sigma\equiv d\sigma/d \Omega$
and $\sigma_0\equiv d\sigma_0/d\Omega$) can be divided into three
classes of double polarization experiments: {\bf
\begin{itemize}
\item polarized photons and polarized target\vspace{-5mm}
\end{itemize}
}
\be
  \sigma  =
      \sigma_0 \bigl[\, 1 - P_{-}^{\gamma} \Sigma \cos 2 \Phi
+ P_{x}^t ( - P_{-}^{\gamma} H \sin 2 \Phi + P_{\odot}^{\gamma} F )
      \nonumber \\
\phantom{t}\hspace*{-4mm} - P_{y}^t ( - T + P_{-}^{\gamma} P \cos 2
\Phi ) - P_{z}^t ( - P_{-}^{\gamma} G \sin 2 \Phi
-P_{\odot}^{\gamma} E )
      \bigr]\hspace*{4mm}
\label{beam_target}
\ee
which reduces to
\be
\sigma &=&
\sigma_0\,\{\,1 +\, P_{\,\odot}^{\gamma}\,\,P_{z}^t\, E \nonumber\\
&-&P_{-}^{\gamma}\,\Sigma\,\cos\,2\Phi\,+
\,P_{-}^{\gamma}\,P_{z}^t\, G\,\sin\,2\Phi \,\}
\label{equation2}
\ee
 for target polarization $\vec P^t=(0,0,P_{z}^t)$ in beam
direction. Here we changed the sign of $E$ polarization compared to
\cite{maid} to be in correspondence with other analyses (e.g.
\cite{Barker:1975bp}).

\begin{figure}[h!]
\vspace*{-1cm} \centerline{
\epsfig{file=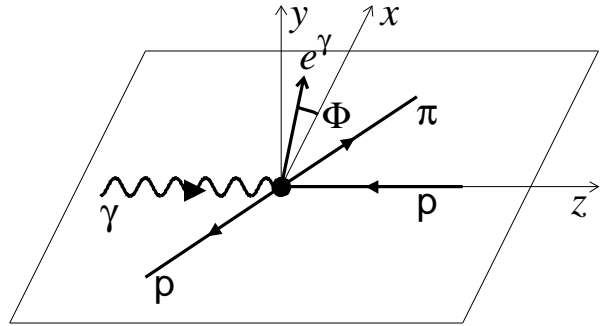,width=0.47\textwidth,clip=} }
\caption{\label{maid_plane} The definition of the $\Phi$ angle for
the single-pion photoproduction, (from \cite{maid})}
\end{figure}

{\bf
\begin{itemize}
\item polarized photons and recoil polarization \vspace*{-5mm}
\end{itemize}
}
\begin{eqnarray}
  \sigma  =
      \sigma_0 \bigl[\, 1 - P_{-}^{\gamma} \Sigma \cos 2 \Phi
      + P_{x}^r ( - P_{-}^{\gamma} O_{x'} \sin 2 \Phi -
P_{\odot}^{\gamma} C_{x'})      \nonumber \\
\phantom{t}\hspace{-4mm} - P_{y}^r ( - P + P_{-}^{\gamma} T \cos 2
\Phi )
      - P_{z}^r ( P_{-}^{\gamma} O_{z'} \sin 2 \Phi +
P_{\odot}^{\gamma} C_{z'}) \bigr]  \hspace*{4mm} \end{eqnarray}
{\bf
\begin{itemize}
\item polarized target and recoil polarization\vspace*{-5mm}
\end{itemize}
}
\begin{eqnarray}
  \sigma  =
      \sigma_0 \bigl[\, 1 + P_{y}^r P
      + P_{x}^t ( P_{x}^r T_{x'} + P_{z}^r T_{z'}) \nonumber \\
 + P_{y}^t ( T + P_{y}^r \Sigma )
      - P_{z}^t \left( P_{x}^r L_{x'} - P_{z}^r L_{z'} \right)
      \bigr]\,.
\end{eqnarray}
In all cases $\Phi$ denotes the angle between the photon
polarization vec\-tor and the reaction plane ($xz$) (see
Fig.\ref{maid_plane}). The polarization of the recoil ba\-ryon is
measured in a coordinate system $(x'\!,y',z')$ where $z'$ is defined
by the direction of the recoil baryon, $y'=y$ and $x'$ is orthogonal
to $y',z'$.

\subsection{Coordinate frame fixed by the polarization of initial particles}

\begin{figure}[h!]
\vspace*{-1cm} \centerline{
\epsfig{file=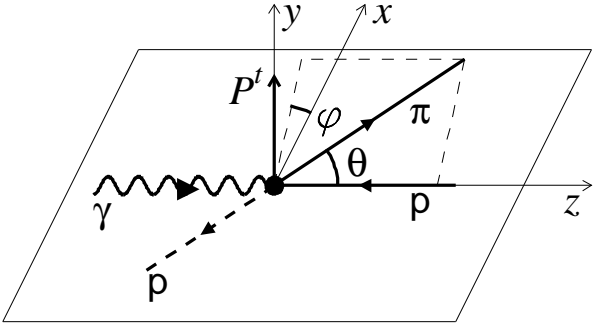,width=0.47\textwidth,clip=} }
\caption{\label{varphi_target} The definition of the $\varphi$ angle
for the single pion photoproduction with the target polarization
directed along the $y$-axis}
\end{figure}

In most papers devoted to polarization variables the amplitude is
calculated in the frame where the $x$-axis together with the
$z$-axis forms the reaction plane. The azimuthal angle $\Phi$ is
defined as the angle between the reaction plane and the direction of
photon polarization. For double polarization observables, e.q.
beam-transverse target, the calculations are made with target
polarization directed along one of the axes in this coordinate
frame. Experimentally, the angle between beam and transverse target
polarization is fixed and it may be advantageous  to fix the axes by
these polarizations.  To extract the dependence of the cross section
on the azimuthal angle in such a frame, eq. (\ref{beam_target}) must
be rotated by $\varphi$ which is angle between the $x$ axis and the
direction of the meson momentum. For unpolarized photons, in the
frame where target polarization (100\%) is directed along $y$-axis:
\be
\sigma  = \sigma_0 \bigl[\, 1 + T \cos \varphi \bigr]
\ee
The definition of the angles is shown in Fig.\ref{varphi_target}.

Let us list here the final expressions for the beam target
polarization experiment. If the target polarization is directed
along the $y$-axis then one obtains for the beam polarizations
directed along $x$ ($\sigma^y_{x}$),  $y$ ($\sigma^y_{y}$), and for
unpolarized beam ($\sigma^y_0$):
\be
&&\frac{\sigma^y_x}{\sigma_0}=1\!-\!\Sigma\cos
2\varphi\!+\!(T\!-\!\frac{H\!+\!P}{2})\cos\varphi\!+\!\frac{H\!-\!P}{2}\cos
3\varphi \nn &&\frac{\sigma^y_y}{\sigma_0}=1\!+\!\Sigma\cos
2\varphi\!+\!(T\!+\!\frac{H\!+\!P}{2})\cos\varphi\!-\!\frac{H\!-\!P}{2}\cos
3\varphi \nn &&\frac{\sigma^y_0}{\sigma_0}=1\!+\!T\cos\varphi\,.
\ee
If the photon polarization is directed along the $x$ axis
$\varphi=\Phi$ from eqs.(\ref{beam_target}) and if it is directed
along the $y$ axis $\varphi=90^o-\Phi$.

With one receives circular polarized beam with helicity $+1$:
\be
\frac{\sigma_\odot^y}{\sigma_0}=1+T\cos\varphi+F\sin\varphi\,.
\label{fpol}
\ee

If the target polarization is directed along $x$-axis we have
\be
\frac{\sigma^x_x}{\sigma_0}&=&1\!-\!\Sigma\cos
2\varphi\!-\!(T\!+\!\frac{H\!+\!P}{2})\sin\varphi\!-\!\frac{H\!-\!P}{2}\sin
3\varphi \nn \frac{\sigma^x_y}{\sigma_0}&=&1\!+\!\Sigma\cos
2\varphi\!-\!(T\!-\!\frac{H\!+\!P}{2})\sin\varphi\!+\!\frac{H\!-\!P}{2}\sin
3\varphi \nn \frac{\sigma^x_0}{\sigma_0}&=&1\!-\!T\sin\varphi\,,
\ee
and with circular polarization (helicity $+1$):
\be
\frac{\sigma_\odot^x}{\sigma_0}=1-T\sin\varphi+F\cos\varphi
\ee
For a photon polarization directed at 45 degrees relative to the
$x$-axis
\be
\frac{\sigma_{45}^0}{\sigma_0}&=& 1-\Sigma\sin 2\varphi
-T\sin\varphi \nn
&+&\frac{H+P}{2}\cos\varphi+\frac{H-P}{2}\cos3\varphi
\ee
is obtained. In all expressions a 100\% beam and target polarization
is assumed.

It is seen that the extraction of the polarization variables is a
difficult task: e.g. if the target polarization is directed along
the $y$-axis and the beam polarization along the $x$-axis, the
polarized cross section should be first to be decomposed into three
$\varphi$ dependencies which provide the beam asymmetry and two
combinations of  $T$, $H$ and $P$ observables. Due to orthogonality
condition
\be
&&\int\limits_{0}^{2\pi}\cos (n\varphi)\,\frac{d\varphi} {2\pi}=
\int\limits_{0}^{2\pi}\sin(n\varphi)\,\frac{d\varphi} {2\pi}=0 \nn
&&\int\limits_{0}^{2\pi}\cos^2(n\varphi)\,\frac{d\varphi} {2\pi}=
\int\limits_{0}^{2\pi}\sin^2 (n\varphi)\,\frac{d\varphi}
{2\pi}=\frac 12
\label{phi_orthog}
\ee
for any $n=1,2,3,\ldots$, the $T-(H+P)/2$ combination can be
extracted directly from the polarization data by weighting the
events with $\cos\varphi$:
\be
T-\frac{H+P}{2}\!=\!\frac{2}{\sigma_0}\int\limits_{0}^{2\pi}\sigma_{x}^y
\cos \varphi\,\frac{d\varphi} {2\pi}
\ee
Such variable can also be directly compared with theoretical
prediction.

\subsection{Two pseudoscalar meson photoproduction}

For the photoproduction of two mesons the number of polarization
variables is much larger. Already in the case of polarized beam and
unpolarized target a beam asymmetry can be extracted for every
outgoing particle. Moreover, the polarization observables can be
extracted from the $\varphi$ dependencies calculated in the
center-of-mass system of two final particles. There are also new
observables compared to single meson photoproduction. For example,
in the case of the single meson photoproduction the circular
polarization of the beam does not provide any additional information
if the target is unpolarized and the polarization of the final
proton is not measured. However in two meson photoproduction the
circular polarization alone can effectively provide a beam-recoil
asymmetry: for example in the case of $\Delta\pi$ final state some
information about $\Delta$ polarization can be obtained from its
decay into $\pi N$.

\subsection{Vector meson photoproduction}

Polarization variables have been extensively studied  in $\omega$
photoproduction for the case of the linear polarized beam and
unpolarized target and, in particular, for the $\omega$ meson
decaying into three pions, e.g
\cite{Schilling:1969um,Zhao:2000tb,TL02}. For linear (or circular)
polarized beam and polarized target, and radiative omega decay
into $\pi^0\gamma$ the theory is not fully developed. In the present
paper we concentrate on polarization observables in the
photoproduction of $\omega$ mesons with subsequent decays $\omega\to
\pi^0\gamma$. The formalism is the same for the radiative $\Phi$
decays into $\eta \gamma$ (with a branching ratio of $BR=1.3\%$)
and $\pi^0\gamma$ ($BR=0.13\%$).

\begin{figure}[h!]
\vspace*{-1cm} \centerline{
\epsfig{file=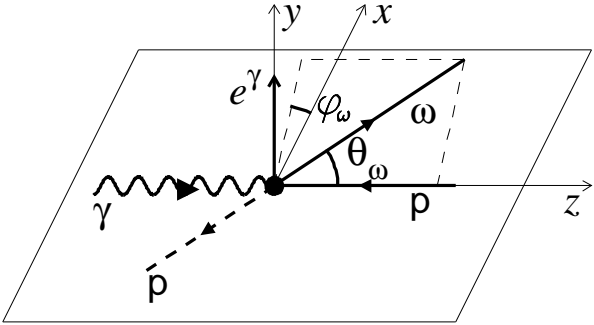,width=0.47\textwidth,clip=} }
\caption{\label{varphi_omega} The definition of the $\varphi_\omega$
angle for the $\omega$-meson photoproduction with the beam
polarization directed along the $y$-axis}
\end{figure}

The $z$-axis is usually chosen along the initial beam. For
unpolarized single pseudoscalar-meson photoproduction there is no
additional vector to fix the azimuthal angle. It is a 'one-plane'
reaction and the measured cross section should not depend how this
plane is oriented. In the case of unpolarized vector meson
photoproduction the situation is more complicated: if the
polarization of the $\omega$ is, at least partly, measured, then there
are two planes involved  and dependencies on the relative azimuthal
angle of the $\omega$ and $\pi^0$ can be obtained. This is similar
to the situation with two pseudoscalar mesons in the final state.

With linearly polarized beam the cross section should not change,
if the photon polarization vector is rotated by 180 degrees and, 
therefore, the beam
polarization enters with a $2n\varphi$ dependence. In $\omega$
photoproduction three beam asymmetries can be naturally extracted:
the beam asymmetry of the $\omega$ meson calculated in the c.m.s. of
the reaction, $\Sigma$ (omega asymmetry), the beam asymmetry of the
final pion in the c.m.s. of the reaction, $\Sigma_\pi$ (pion
asymmetry), and the beam asymmetry of the final pion in the rest
frame of the $\omega$, $\Sigma^\omega_\pi$ (pion-in-omega-rest-frame
asymmetry). The decay asymmetry $\Sigma_d$ introduced for the decay
of vector mesons into two and three pseudoscalar particles (see e.g.
\cite{Schilling:1969um,Criegee:1970av,Ball73}) is
extracted as a coefficient in front of the $\cos 2\Psi$ dependence
in the production cross section, 
where $\Psi$ is the angle between the photon polarization and the momentum
of the final pseudoscalar meson. If this definition is extended to
$\omega$'s decaying into $\pi^0\gamma$, the decay asymmetry 
differs by a sign from $\Sigma^\omega_\pi$:
\be
\Sigma_d=-\Sigma^\omega_\pi\,.
\ee

\begin{figure}[h!]
\vspace*{-1cm} \centerline{
\epsfig{file=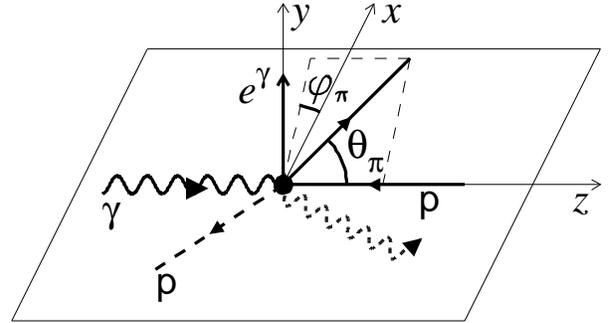,width=0.47\textwidth,clip=} }
\caption{\label{varphi_pi} The definition of the $\varphi_\pi$ angle
for the $\omega$-meson photoproduction with the beam polarization
directed along the $y$-axis}
\end{figure}

By means of the orthogonality condition (\ref{phi_orthog}) the
asymmetries can be defined from the cross section obtained with a
photon beam polarized linearly along the $y$-axis as:
\be
&&\Sigma\!=\!\frac{2}{\sigma_0}\!\int\limits_{0}^{2\pi}\!\sigma_{y}
\cos 2\varphi_\omega\frac{d\varphi_\omega}{2\pi},\quad
\Sigma_\pi\!=\!\frac{2}{\sigma_0}\!\int\limits_{0}^{2\pi}\!\sigma_{y}
\cos 2\varphi_\pi\frac{d\varphi_\pi}{2\pi},\nn
&&\Sigma^\omega_\pi\!=\!\frac{2}{\sigma_0}\!\int\limits_{0}^{2\pi}\!\sigma_{y}
\cos 2\varphi^\omega_\pi\frac{d\varphi^\omega_\pi}{2\pi}\,.
\label{sigma_int}
\ee
The definition of the $\varphi_\omega$ and $\varphi_\pi$ angles are
shown in Figs. \ref{varphi_omega},\ref{varphi_pi}. The beam
asymmetries can be extracted directly from the experimental data
through their specific $\varphi$ dependence.

This also holds for the beam-target double polarization observables,
e.g of G-type. The $G$, $G_\pi$, $G^\omega_\pi$ observables can be
calculated from the polarized cross section obtained with linear
polarized beam (e.g. along the $y$-axis) and target polarization
along the $z$-axis. These variables have a $\sin 2\varphi$
dependence:
\be
&&G\!=\!\frac{2}{\sigma_0}\!\int\limits_{0}^{2\pi}\!\sigma_{y} \sin
2\varphi_\omega\frac{d\varphi_\omega}{2\pi},\quad
G_\pi\!=\!\frac{2}{\sigma_0}\!\int\limits_{0}^{2\pi}\!\sigma_{y}
\sin 2\varphi_\pi\frac{d\varphi_\pi}{2\pi},\nn
&&G^\omega_\pi\!=\!\frac{2}{\sigma_0}\!\int\limits_{0}^{2\pi}\!\sigma_{y}
\sin 2\varphi^\omega_\pi\frac{d\varphi^\omega_\pi}{2\pi}\,\,
\label{g_int}
\ee
The observable $E$ is the asymmetry of helicity $3/2$ and $1/2$
initial states,
\be
E=\frac{\sigma_{3/2}-\sigma_{1/2}}
{\sigma_{3/2}+\sigma_{1/2}}\,.
\label{e_int}
\ee
It is measured using circular polarized beam and longitudinal
polarized target. There is no dependence on any $\varphi$ angle
additional to the unpolarized case.

\section{Sensitivity of the polarization variables to
the vectormeson production dynamics}

The main contribution to the $\gamma p\to \omega p$ cross section at
photon energies above 3\,GeV  comes from the fluctuation of the photon
into a vector meson. In this regime the cross section exhibits
diffractive behavior \cite{Ball70,Ball73,Lieb83,Atki84} and has an
exponential $t$--dependence
\begin{equation} \frac{d\sigma}{dt} \quad \simeq \quad
\frac{d\sigma}{dt}{\Bigg\vert}_{t=0} \cdot e^{-b|t|},
\end{equation}
where the parameter $|b|$ characterizes the range of the
vector-meson-nucleon interaction. The photon helicity is transferred
to the vector meson, i.e. the $s$--channel helicity is conserved.
This requires $(-)^J$ natural parity of the exchange particle, i.e.
$0^+$, $1^-$, etc. (``Pomeron''), and intricate conspiracy between
$t$--channel spin--flip and non spin--flip contributions
\cite{Gilman70}. At energies below 3\,GeV both exchange--parity
characteristics are expected to contribute on a similar level. The
unnatural parity admixture is attributed to $t$--channel exchange of
pseudoscalar mesons, in particular $\pi^0$ \cite{FS96}. Resonances
in the $s$--channel are preferably expected to show up in the
threshold region, since they would be form--factor suppressed at
higher energies.

Polarization observables have been extensively studied in $\omega$
photoproduction for the case of linear polarized beam and
unpolarized target and, in particular, for the $\omega$ decaying into
three pions (see e.g. \cite{Schilling:1969um}).
For forthcoming experiments with linearly or circularly polarized beam and
polarized target which use the radiative  $\omega$ decay
the theory is not fully developed.
In the following we examin characteristic signatures of the different
pieces of the reaction mechanism on the polarization observables in the
reaction $\gamma\,p \to \omega\,p \to \gamma\pi^0\,p$.

\subsection{Pion exchange}

The $\omega$ meson decays in a P-wave into the $\pi^0\gamma $ system.
The amplitude is described by the convolution of the antisymmetrical
tensor with the momentum of the photon, the momentum of the
$\omega$-meson and the polarization vectors of omega and photon.

In the case of a pion exchange in the $t$-channel the omega
production is described by the amplitude
\be
&&\tilde A^{ij}_{\gamma p\to \omega p}=A^{ij}_{\pi^*} A(p\to
\pi^*p),\,\,\,{\rm where} \nn &&A^{ij}_{\pi^*}= \epsilon^{(i)}_\mu
\varepsilon_{\mu\alpha k p} \omega_\alpha= \epsilon^{(i)}_\mu
\varepsilon_{\mu\alpha k p}\frac{1}{M_\omega} \varepsilon_{\nu\alpha
q p}\, \tilde \epsilon^{(j)}_\nu \,.
\label{pion}
\ee
Here $k$ is the momentum of the initial photon ($k\equiv k_2$), $q$
the momentum of the final pion ($q\equiv q_2$), $p$  the momentum of
the $\omega$-meson and
\be
\varepsilon_{\mu\alpha q p}\equiv \varepsilon_{\mu\alpha \mu\nu}
q_\mu p_\nu.
\ee
The $\epsilon^{(i)}_\mu$ and $\tilde \epsilon^{(j)}_\nu$ are the
polarization vectors of the initial and final photon, respectively.
The amplitude $A(p\to \pi^*p)$ describes the propagation of the
virtual pion ($\pi^*$) and its absorption by the proton.

The unpolarized cross section is proportional to the squared
amplitude summed over the polarization of the final photon and
nucleon, and averaged over the polarization of the initial photon
and nucleon:
\be
\sigma_0=\frac 14 \sum\limits_{i,j=1}^2|A^{ij}_{\pi^*}|^2
\frac{(2\pi)^4|A(p\to \pi^*p)|^2}{4\sqrt{(k_1k_2)^2-m_1^2m_2^2}}
d\Phi_n
\ee
where $d\Phi_n$  is defined by eq.(\ref{phv}). The amplitude for the
interaction of the virtual pion with the nucleon can be factorized
out from the production of the $\omega$-meson.

We first consider the  beam asymmetry in the rest frame of the omega
meson. If the $z$-axis is directed along the beam, then:
\be
q=(q_0;|\vec q|\sin\Theta_\pi\cos\varphi_\pi, |\vec
q|\sin\Theta_\pi\sin\varphi,|\vec q| \cos\Theta_\pi) \nn
p=(M_\omega;0,0,0)\qquad \qquad k=(k_0;0,0,k_z)\,.~~~~
\ee

The $y,x$-axes can be chosen in the way that two orthogonal
polarization vectors of the initial photon (which are
$\epsilon^{(y)}=(0;0,1,0)$ and $\epsilon^{(x)}=(0;1,0,0)$ in c.m.s.
of the reaction) have also only one component in the $xy$ plane in
the omega rest frame. The boost of these vectors to the $\omega$
rest frame yields:
\be
\epsilon^{(y)}&=&(-\frac{\omega_y}{M_\omega};0,1,-\frac{\omega_y}{M_\omega})\,,
\nn
\epsilon^{(x)}&=&(-\frac{\omega_x}{M_\omega};1,0,-\frac{\omega_x}{M_\omega})
\label{e_orf}
\ee
where $\omega_x$ and $\omega_y$ are the $x$ and $y$-components of
the $\omega$-meson momentum in c.m.s. of the reaction:
\be
p=(\omega_0;\omega_x,\omega_y,\omega_z) \qquad {\rm in\,\, c.m.s.}.
\label{omega_mom_cms}
\ee

Summing over the polarization of the final photons (of  eq.
(\ref{omega_fin}) we obtain for the squared amplitude:
\be
|A^i|^2=\sum\limits_{j=1}^2 |A^{ij}|^2=-\epsilon^{(i)}_\mu
\epsilon^{(i)*}_\nu \varepsilon_{\mu\alpha k p}
\varepsilon_{\nu\beta k p} \times \nn \left (q_\perp^2
g^\perp_{\alpha\beta}-q^\perp_\alpha q^\perp_\beta\right )\,.~~
\label{a2_pion}
\ee
Due to the fact that the omega 4-momentum has only an energy
component and the momentum of the initial photon has only energy and
$z$ components, the antisymmetrical tensor can be written as
\be
\varepsilon_{\mu\alpha k p}=\varepsilon_{\mu\alpha z 0} k_z
M_\omega\,.
\ee
Therefore only  $y$ and $x$ components of the polarization vectors
of the initial photon give a contribution to the cross section. For
the squared amplitude we then obtain:
\be
|A^y|^2&=&k^2_z M^2_\omega \left (|\vec q|^2-q_x q_x\right ) \nn
|A^x|^2&=&k^2_z M^2_\omega \left (|\vec q|^2-q_y q_y\right )\,.
\ee
Averaging over the polarization of the initial photon yields
\be
\frac{|A^x|^2+|A^y|^2}{2}&=& k^2_z M^2_\omega \left (\vec
q^2-\frac{q_x q_x+q_y q_y}{2}\right ) \nn&=& k^2_z
M^2_\omega\frac{\vec q^2}{2}(1+z^{\omega\,2}_\pi)\,,
\label{unpol_pion}
\ee
where $z^\omega_\pi\equiv \cos \Theta_\pi$ is defined in the rest
frame of the $\omega$-meson.

If the beam polarization vector is chosen along the $x$-axis, then
\be
&&|A^x|^2=\frac{|A^x|^2+|A^y|^2}{2}+k^2_z M^2_\omega \frac{q_x
q_x-q_y q_y}{2}=\nn && \frac{|A^x|^2+|A^y|^2}{2}+k^2_z M^2_\omega
\frac{|\vec q|^2}{2}(1-z^{\omega\,2}_\pi)\cos 2\varphi_\pi^\omega
\ee
is obtained. Assuming 100\% polarization:
\be
\sigma_x=\sigma_0(1-\Sigma^\omega_\pi\cos 2\varphi_\pi^\omega),
\ee
and we obtain:
\be
\Sigma^\omega_\pi=-\frac{1-z^{\omega\,2}_\pi}{1+z^{\omega\,2}_\pi}\,.
\label{pion_zpi}
\ee
If the photon polarization is directed along the $y$ axis
\be
\sigma_y=\sigma_0(1+\Sigma^\omega_\pi\cos 2\varphi_\pi^\omega)\,.
\ee

\begin{figure}[h!]
\centerline{ \epsfig{file=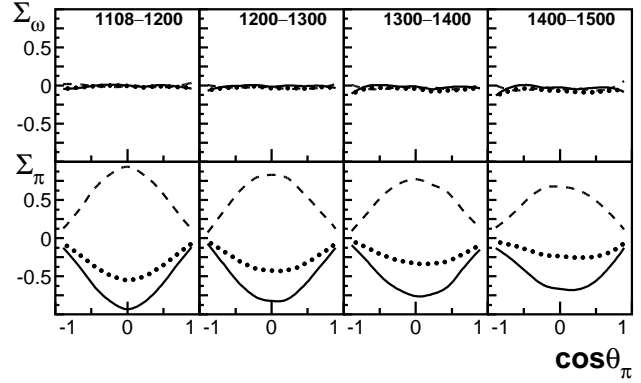,width=0.47\textwidth,clip=} }
\caption{\label{cms_zpi} Prediction for the omega asymmetry
$\Sigma_\omega$ and the pion asymmetry $\Sigma_\pi$ as a function
of the angle between beam and $\pi$-meson calculated in the c.m.s.
of the reaction in the indicated bins of photon energy between
threshold and $E_\gamma=1500$ MeV. The asymmetry related to pion
exchange is shown as solid line, pomeron exchange as dashed
line and mixture (see text) of pion and pomeron exchanges
as dotted line.}
\end{figure}

There is no dependence of $\Sigma^\omega_\pi$ on the other angles,
e.g. the angle of the omega in the c.m.s. of the reaction. Thus,
integrating the polarized and non-polarized cross section over
$z^\omega_\pi$ we obtain in the omega rest frame.
\be
\Sigma^\omega_\pi=-\frac{1-1/3}{1+1/3}=-\frac 12\,.
\label{pion_zom}
\ee

\begin{figure}[h!]
\centerline{ \epsfig{file=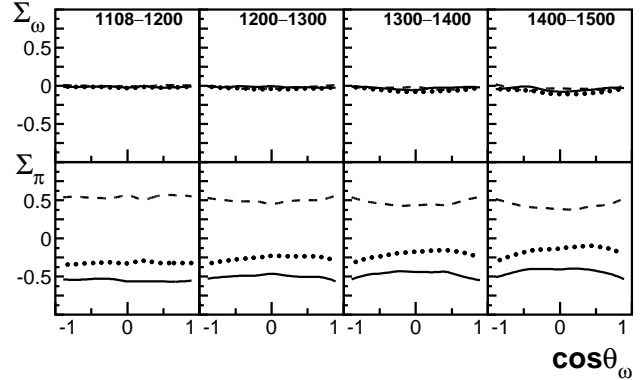,width=0.47\textwidth,clip=} }
\caption{\label{cms_omega} Prediction for the omega asymmetry
$\Sigma_\omega$ and the pion asymmetry $\Sigma_\pi$ as a function
of the angle between beam and $\omega$-meson calculated in the
c.m.s. of the reaction.
Notation of curves and energy bins is the same as in
Fig.~\ref{cms_zpi}.}
\end{figure}

In the non-relativistic limit where $|\vec \omega|<M_\omega$
eq.(\ref{omega_mom_cms}) the boost from the $\omega$ rest frame to
the c.m.s. frame does not change the polarization vectors of the
photon. This means that the dependence of $\Sigma_\pi$ on the
$z_\pi$ (calculated in c.m.s. of the initial photon and proton)
should be the same as the dependence of $\Sigma^\omega_\pi$ on
$z^\omega_\pi$ calculated in the rest frame of omega. With
increasing energy the boost becomes more important.  To extract
these variables we performed simulations of $\Sigma_\pi$ and
$\Sigma_\omega$ with the Bonn-Gatchina partial wave analysis
program. Here the pion exchange amplitude was taken in a reggeized
form and the polarized and unpolarized cross sections were
calculated from a sample of 400 000 $\gamma p\to \omega p$ Monte
Carlo events. The angular dependence of $\Sigma_\pi$ and
$\Sigma_\omega$ on $z_\pi$ at different energies of the initial
photon is shown in Fig.\ref{cms_zpi}. $\Sigma_\pi$ has the angular
dependence close to $(1-z^2_\pi)/(1+z^2_\pi)$ while the omega
asymmetry is equal to zero. Indeed,  for pure $\pi^*$ exchange the
cross section measured with linear polarized beam carries no
dependence on the azimuthal angle of the $\omega$-meson and hence
this asymmetry is expected to vanish. The dependence of the
$\Sigma_\pi$ and $\Sigma_\omega$ observables on the $\cos
\Theta_\omega$ is shown in Fig.\ref{cms_omega}. Both variables show
no dependence on this angle at small energies. Again, as expected,
the $\Sigma_\pi$ asymmetry is close to $-0.5$ and $\Sigma_\omega$ is
equal to zero. Only at large energies a weak dependence of
$\sigma_\pi$ on $\cos \Theta_\omega$ is observed.

\subsection{Pomeron exchange}

To investigate the pomeron exchange the leading
$\gamma$-Pomeron-$\omega$ vertex with S-wave orbital moment between
pomeron and initial photon is considered. The omega photoproduction
amplitude for such a mechanism has the form:
\be
&&\tilde A^{ij}_{\gamma p\to \omega p}=A^{ij}_{0^+} A(p\to 0^+p) \nn
&&A^{ij}_{0^+}= \epsilon^{(i)}_\mu \omega_\mu=\epsilon^{(i)}_\mu
\frac{1}{M_\omega}\varepsilon_{\nu\mu q p}
\,\tilde\epsilon^{(j)}_\nu \,.
\label{pom}
\ee
Here $A(p\to 0^+p)$ is the amplitude describing the pomeron
propagation and the interaction of the pomeron with the nucleon.
Summing over the polarizations of the final photons yields:
\be
|A^i_{0^+}|^2=\sum\limits_{j=1}^2
|A^{ij}_{0^+}|^2=-\epsilon^{(i)}_\mu \epsilon^{(i)*}_\nu \left
(q_\perp^2 g^\perp_{\mu\nu}-q^\perp_\mu q^\perp_\nu\right )\,.~~
\label{a2_pom}
\ee
Here the z-component of the polarization vector is to be taken into
account. Then
\be
|A^y_{0^+}|^2&=& |\vec q|^2(1+\frac{w_y^2}{M_w^2})-(q_y
-q_z\frac{w_y}{M_\omega})^2 \quad {\rm and}\nn |A^x_{0^+}|^2&=&
|\vec q|^2(1+\frac{w_x^2}{M_\omega^2})-(q_x
-q_z\frac{w_x}{M_\omega})^2
\ee
The unpolarized cross section is proportional to:
\be
&&\frac{|A^x|^2\!+\!|A^y|^2}{2}= \frac{|\vec q|^2}{2}\Big
(1+z_\pi^{\omega\,2}\!+\! \frac{|\vec
w|^2}{M_\omega^2}(1\!-z_\pi^{\omega\,2})(1\!-z_\omega^2) \nn
&&+2\frac{z^\omega_\pi|\vec
w|}{M_\omega}\sqrt{(1\!-\!z_\pi^{\omega\,2})(1\!-\!z_\omega^2)}
\cos(\varphi^\omega_\pi-\varphi_\omega)\Big ).~~~
\ee
There is a dependence on the difference between the angles
$\varphi_\omega$ and $\varphi^\omega_\pi$. This is a consequence of
the fact that the polarization vector of the $\omega$-meson is
partially reconstructed through the $\gamma\pi$ decay. In a sense we
dial with an "incomplete" recoil polarization experiment.

\begin{figure}[h!]
\centerline{
\epsfig{file=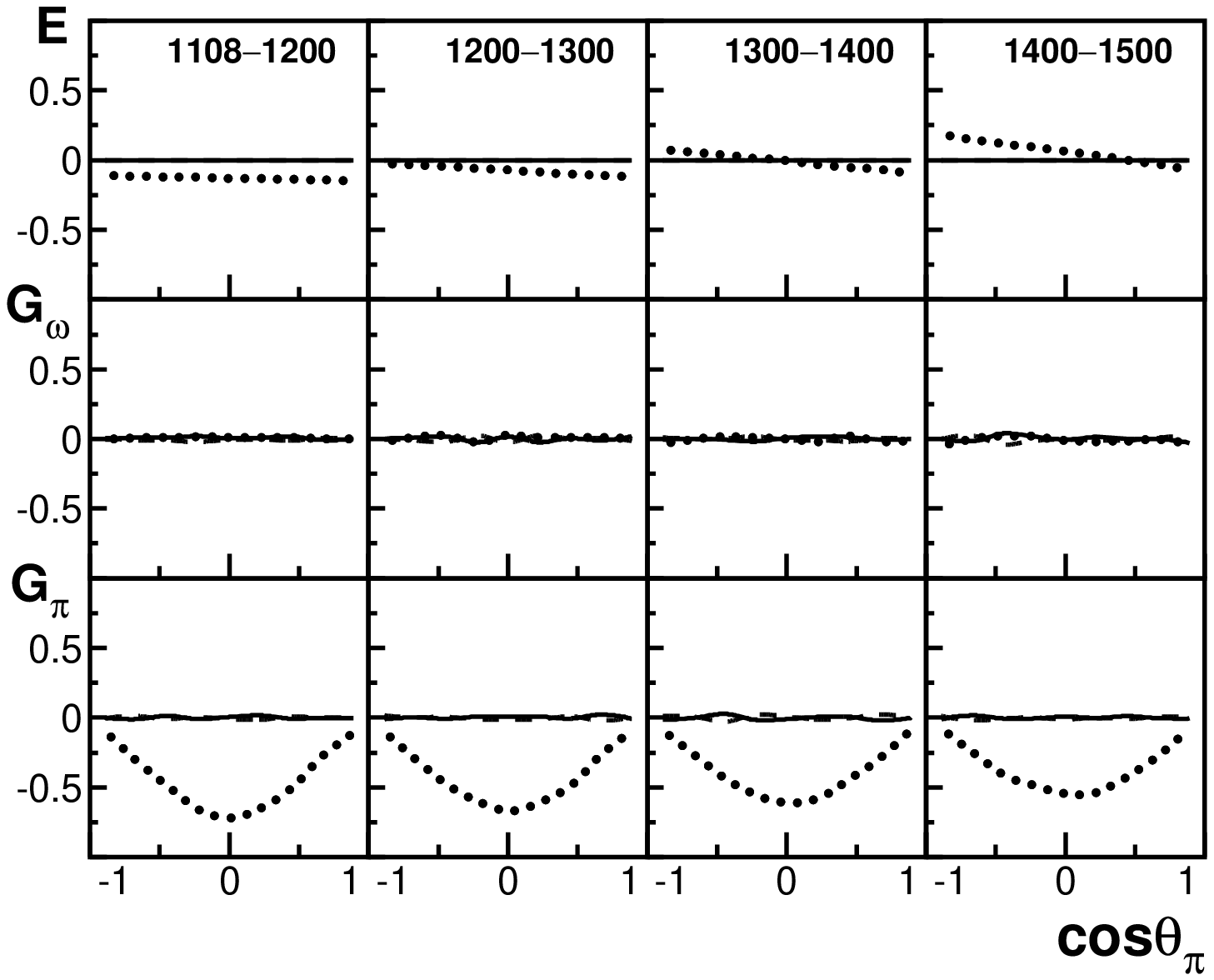,width=0.47\textwidth,clip=} }
\caption{\label{eg_pi} Prediction for the helicity asymmetry
E and the asymmetries $G$ and $G_\pi$  as a function of the angle
between beam and $\pi$-meson calculated in the c.m.s. of the
reaction.
Notation of curves and energy bins is the same as in
Fig.~\ref{cms_zpi}.}
\end{figure}

\begin{figure}[h!]
\centerline{ \epsfig{file=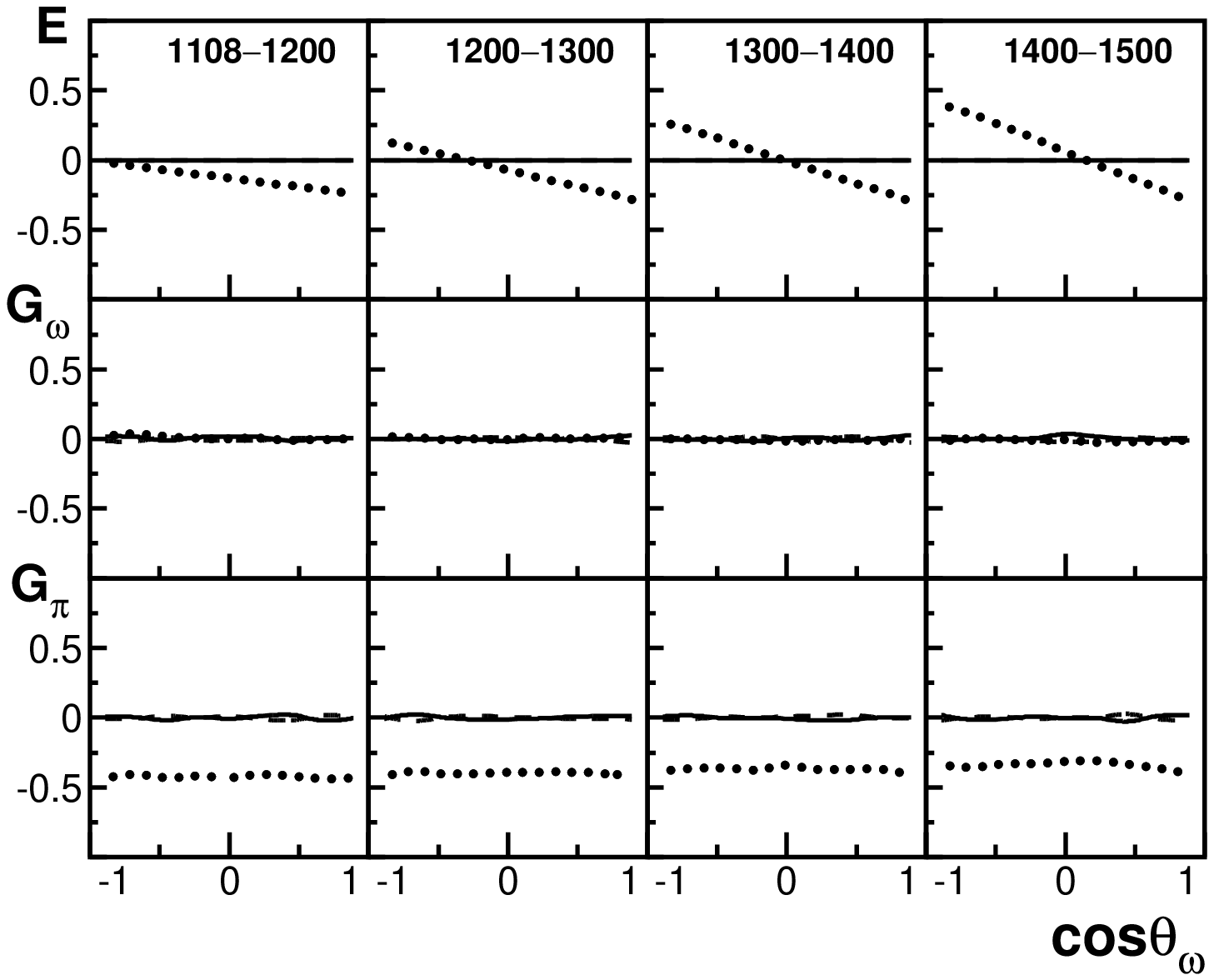,width=0.47\textwidth,clip=}
} \caption{\label{eg_omega} Prediction of the helicity asymmetry E and
the asymmetries $G$ and $G_\pi$  as a function of the angle between
beam and $\omega$-meson calculated in the c.m.s. of the reaction.
Notation of curves and energy bins is the same as
in Fig.~\ref{cms_zpi}.}
\end{figure}

For the initial photon polarization vector along the $x$-axis one
receives:
\be
|A^x|^2&=&\frac{|A^x|^2+|A^y|^2}{2}-\frac{|\vec q|^2}{2}\Big (\cos
2\varphi^\omega_\pi (1-z_\pi^{\omega\,2}) \nn &-& \cos
2\varphi_\omega \frac{|\vec
w|^2}{M_\omega^2}(1\!-\!z_\pi^{\omega\,2})(1\!-\!z_\omega^2)\nn
&-&2\frac{z^\omega_\pi|\vec
w|}{M_\omega}\sqrt{(1\!-\!z_\pi^{\omega\,2})(1\!-\!z_\omega^2)}
\cos(\varphi^\omega_\pi+\varphi_\omega)\Big ).~~~~
\ee
Extracting the $\cos 2\varphi^\omega_\pi$ dependence only and
integrating over all other $\varphi$ dependencies we obtain
\be
\Sigma^\omega_\pi=\frac{1-z_\pi^{\omega\,2}}{1+z_\pi^{\omega\,2}+
\frac{|\vec w|^2}{M_\omega^2}(1-z_\pi^{\omega\,2})(1-z_\omega^2)}\,.
\ee
At small initial energies where the omega momentum is small against
its energy ($|\vec w|<M_\omega$) this yields:
\be
\Sigma^\omega_\pi=\frac{1-z_\pi^{\omega\,2}}{1+z_\pi^{\omega\,2}}\,.
\label{pom_zpi}
\ee
Similar to the case of pure $\pi$-exchange, this observable does not
depend on the $\omega$-meson production angle. However the sign of
$\Sigma^\omega_\pi$ changes for Pomeron exchange:
\be
\Sigma^\omega_\pi=\frac{1-1/3}{1+1/3}=\frac 12\,
\label{pom_zom}
\ee
Apart from the overall sign, the behavior of the pion asymmetry
$\Sigma_\pi$ is very similar for pure pion and pomeron exchange
mechanisms. The omega asymmetry $\Sigma_\omega$ is zero in both
cases.

Results of simulations of the omega and pion asymmetries using the
Bonn-Gatchina fitting program are shown in Figs.~\ref{cms_zpi} and
\ref{cms_omega} for pure pomeron exchange. As expected $\Sigma_\pi$
hardly deviates from the $(1-z_\pi^2)/(1+z_\pi^2)$ in the threshold
region and $\Sigma_\omega$ vanishes.

\subsection{Mixing of pion and pomeron exchange amplitudes}

To fix the mixing between the pion and pomeron amplitudes we fitted
the unpolarized differential cross section measured by the SAPHIR
collaboration from the $\omega$ production threshold at 1723\,MeV to
2.4\,GeV \cite{Barth03}. The fit roughly reproduced the total cross
section. The description of the differential cross section was
rather limited. The contributions of the two mechanisms to the total
cross section was found to be about 40\% for pomeron and 60\% for
pion exchange.
The results in the beam asymmetries
for such a mixing of the pion and pomeron
exchange amplitudes are shown
as dotted curves in Figs.\,\ref{cms_zpi} and \ref{cms_omega}.
These amplitudes do not interfere if
the nucleon is not polarized. Hence, the omega asymmetry is close to
zero and the $\Sigma_\pi$ asymmetry only weakly depends on the angle
of the omega meson with respect to the beam.

\subsection{Double polarization observables}

\begin{figure}
\centerline{ \epsfig{file=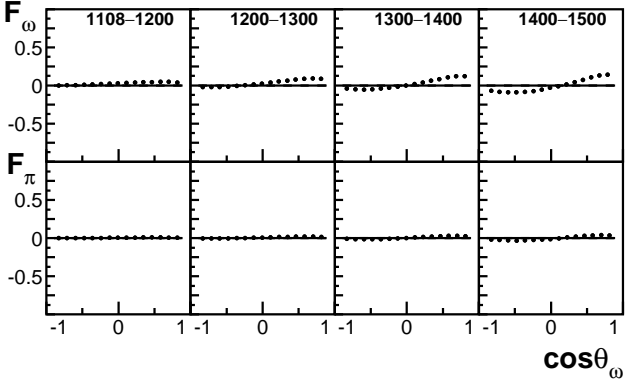,width=0.47\textwidth,clip=}
} \caption{\label{f_omega} Prediction for the $F_\omega$ and $F_\pi$
asymmetries as a function of the angle between beam and
$\omega$-meson calculated in the c.m.s. of the reaction.
Notation of curves and energy bins is the same as
in Fig.~\ref{cms_zpi}.}
\end{figure}

In experiments with polarized beam and polarized target information
about further polarization variables can be extracted. Predictions
for the angular dependence of $E$, $G$ and $G_\pi$ are shown in
Figs.\,\ref{eg_pi} and \ref{eg_omega} for the cases of pion and/or pomeron
exchanges. It can be seen that for pure pion or pomeron exchange all
these double polarization asymmetries are equal to zero. In the case of mixing
of pomeron and pion exchange amplitudes $G_\omega$ remains zero.
However, $E$ shows a clear linear dependence of both $z_\omega$
and $z_\pi$. Also $G_\pi$ exhibits a clear deviation from zero. It
depends on $z_\pi$ like $-A(1-z^2_\pi)/(1+z^2_\pi)$ ($1>A>0$) and is
approximately flat in $z_\omega$.

With transverse target polarization the target asymmetry $T$ is
equal to zero for all three cases. The observables $F_\omega$,
$F_\pi$ can be extracted from measurements with circular polarized
beam. As it is shown in Fig.~\ref{f_omega} they also remain very
small in the case of pion and/or pomeron exchange.

\subsection{Sensitivity of the polarization variables to
contributions from s-channel resonances}
\begin{figure}
\centerline{
\epsfig{file=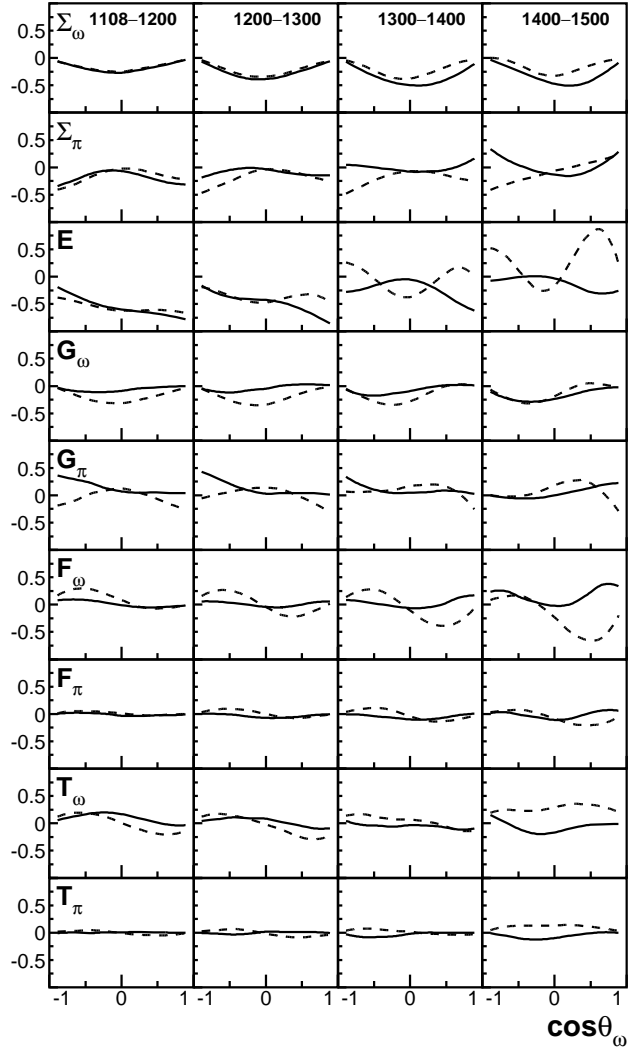,width=0.47\textwidth,clip=} }
\caption{\label{sol_omega} Prediction for the $\cos\Theta_\omega$
dependence of polarization variables for two solutions with dominant
contributions from s-channel resonances. The solution 1 shown by
solid curve and solution 2 by dashed curves.}
\end{figure}

Recent experiments \cite{Barth03,Ajaka06,HS07} indicated that, in
addition to $\pi^0$ and pomeron exchange, s-channel resonances play
a significant role in $\omega$ photoproduction in the threshold
region. Therefore we investigated their impact on the polarization
observables for the $\omega\to \pi^0\gamma$ final state using the
frame of the Bonn-Gatchina partial wave analysis (PWA) program. Reasonably
good agreement with the data is obtained for several different
PWA solutions. Two, labeled (1) and (2), reproduce well the differential
cross sections \cite{Barth03} and the omega asymmetry
$\Sigma_\omega$ even though the physical content of both solutions
is rather different. In (1) two $\rm P_{13}$ resonances provide the
most significant contributions while (2) is dominated by $\rm
N(1680)F_{15}$. Further solutions exist and cannot be distinguished
on the basis of the available data.

Fig.~\ref{sol_omega}  shows the results for single and double
polarization variables of the two solutions (1) and (2). Already
data on the pion asymmetry $\Sigma_\pi$, which can be extracted from
an experiment with linearly polarized beam and unpolarized target,
will help to distinguish
these two solutions. Very large differences are observed in the
helicity asymmetry $E$ (circularly polarized beam and longitudinally
polarized target) and in the observable $F_\omega$ (circularly
polarized beam and transverse target polarization).

\section{Summary and Conclusions}

We have studied single and double polarization observables in the
photoproduction of $\omega$-mesons with subsequent radiative decay
$\omega\to \pi^0\gamma$. It is shown that the pion and pomeron
exchange amplitudes have a very specific signature in the omega
asymmetry $\Sigma_\omega$, the pion asymmetry $\Sigma_\pi$, the
pion-in-omega-rest-frame asymmetry $\Sigma_\pi^\omega$ and in the
double polarization asymmetry $G_\pi$. The measurements of these and
further double polarization observables will be very important to
distinguish between the t-channel exchange contributions and
additional $s$-chan\-nel resonance production. In the Bonn-Gatchina
PWA the observables $E$ and $G_\pi$ appear particularly sensitive to
distinguish between different partial wave solutions which describe
the available data on the same level.

\section*{Acknowledgements}
We would like to thank Eberhard Klempt and Alexander Titov 
for useful discussions and remarks.
This work was supported by the Deu\-tsche
Forschungsgemeinschaft within the SFB/TR16,
and the Russian Foundation for Basic Research
(07-02-01196-a).

\end{document}